# Nonlinear Insulator in Complex Oxides

Z. Q. Liu[1,2], D. P. Leusink[1,2], W. M. Lü[1,3], X. Wang[1,2], X. P. Yang[4], K. Gopinadhan[1,3]

A. Annadi[1,2], S. Dhar[1,3], Y. P. Feng[2], H. B. Su[4], G. Xiong[1,3], T. Venkatesan[1,2,3],

Ariando[1,2*]

[1]*Nanocore*, [2]*Department of Physics*, [3]*Department of Electrical and Computer*

*Engineering, National University of Singapore, Singapore*

[4]*Division of Materials Science, Nanyang Technological University, Singapore*

*To whom correspondence should be addressed.
E-mail: ariando@nus.edu.sg

**Here we report a new class of material which we have coined as "nonlinear**

**insulators" that exhibit a reversible electric-field-induced metal-insulator transition.**

**We demonstrate this behaviour for an insulating $LaAlO_3$ thin film in a**

**metal/$LaAlO_3$/Nb-$SrTiO_3$ heterostructure. Our experimental results exclude the**

**possibility of diffusion of the metal electrodes or oxygen vacancies into the $LaAlO_3$**

**layer. Instead, the phenomenon is attributed to the formation of a quasi-conduction band (QCB) in the defect states of LaAlO$_3$ that forms a continuum state with the conduction band of the Nb-SrTiO$_3$. Once this continuum (metallic) state is formed, the state remains stable even when the voltage bias is turned off. An opposing voltage is required to deplete the charges from the defect states.**

The insulating state is one of the most basic electronic phases in condensed matter. This state is characterized by an energy gap for electronic excitations that makes an insulator electrically inert at low energy. However, for complex oxides, the very concept of an insulator must be re-examined. Complex oxides behave differently from conventional insulators such as $SiO_2$, on which the entire semiconductor industry is based, because of the presence of multiple defect levels within their bandgap.

The electronic state of complex oxides can be changed by chemical doping, temperature, external pressure (*1-3*), magnetic fields (*4*), electric fields (*5*) or light (*6*). In particular, the electric field-induced metal-insulator transition (MIT) has attracted a lot of attention because of its intriguing physical mechanisms (*7, 8*) and potential for device applications. With the recent excitement in resistive switching (*9-11*) in a typical metal/insulator/metal structure (*12, 13*), electric-field-induced MIT has been revisited as a possible mechanism (*14-16*). A number of different mechanisms (*17-20*) have been previously demonstrated for resistive switching, such as electric-field-induced oxygen vacancy migration resulting in the formation of conducting filaments (*17, 18*), and reversible metal migration from electrodes (*19*). Such a variety of field-induced phenomena arise from the complex defects that are present in these oxides. Cationic and anionic defects can form trap states within the bandgap of even wide bandgap oxides, drastically affecting their insulating properties (*21*). In this letter, we show a reversible metal-insulator transition caused by the electrical population of defect levels in the bandgap of oxide insulators, which in turn can play a crucial role in determining the (nonlinear) insulating nature of the material. The very large bandgap of $LaAlO_3$, ~5.6 eV (*22*), and the possibility of fabricating high crystalline quality $LaAlO_3$ thin films make it an ideal material for investigation of the nonlinear insulating properties of complex oxides.

Typical metal/$LaAlO_3$/Nb-$SrTiO_3$ structures (Fig. 1A) with different top electrodes (Cu or Au) and different $LaAlO_3$ layer thicknesses of 25, 50 and 150 nm were prepared (supplementary information). Figure 1 shows that the electronic phase of the Cu/$LaAlO_3$ (150 nm)/Nb-$SrTiO_3$ structure can be reversibly changed by the applied voltages. In the following, we will describe the switching sequence starting with the negative voltage. The temperature dependence of the sample resistance (*R-T*) curve of the initial state indicates typical insulating behaviour (Fig. 1B) that corresponds to the initial current-voltage (*I-V*) curve of the system (from i to j in Fig. 1C). At low voltages, the resistance is very high but when the voltage reaches −6.8 V, a sharp jump in the current is seen (Fig. 1C). Because of the imposed compliance current of 0.1 A, which is used to prevent sample damage (*12, 19*), the current saturates at 0.1 A. As the voltage is scanned back to 0 V, a linear *I-V* characteristic is seen, which corresponds to a stable metallic state (inset m in Fig. 1C). Hence, upon the dramatic switching of the resistive state triggered at −6.8 V, the resistance changes from ~14 MΩ to ~25 Ω at room temperature, which is concomitant with a phase transition from the insulating to the metallic phase. The metallic state persists until the positive voltage scan reaches about 2.4 V; at that level, the structure transitions into a high-resistance (n-o-p-q in Fig. 1C) and non-metallic (inset q in Fig. 1C) state. The reversible phase transition is reproducible, as shown in the second *I-V* cycle (Fig. 1D). These behaviours are totally unexpected for a wide bandgap insulator such as $LaAlO_3$.

The anomalous insulating behaviour of $LaAlO_3$ cannot be explained by artefacts such as anodization or redox of the active metal electrodes (*19*). In those cases, anodic dissolution of the metal electrode is possible only if adequate positive voltage is applied to an active metal electrode, and the resulting cations can be driven by the strong electric field into the insulating film where they form metallic filaments. To further eliminate this possibility, structures with inactive Au as the top electrode material were prepared and analysed; these structures showed no observable difference in their resistive switching behaviour compared to those that had Cu as the top electrode (supporting online material). This strongly suggests that the reversible phase transition cannot be caused by the diffusion of metal electrodes.

Another possible mechanism for this anomalous insulating phenomenon is the migration of oxygen vacancies. If such a mechanism is occurring, the applied electric field could only change the distribution of the positively charged oxygen vacancies (*23*). Therefore, no insulating phase would appear, even at higher resistance states. Our case is also significantly distinct from the type of resistive switching that originates from the electro-migration of excess oxygen as described by Shi *et al.* (*24*); in that experiment, the resistive switching disappeared at low temperatures because of the low diffusion coefficient of the oxygen. To verify this, the low temperature switching properties were examined using the same structure with a Au electrode. The initial insulating state of this structure is shown in the inset i of Fig. 2A. After cooling to 4.1 K, the *I-V* measurements were performed and the structure was negatively biased to −8.8 V, at which it switched to a linear state (Fig. 2A) with a very small resistance of ~12 Ω, which corresponds to a metallic phase (inset n in Fig. 2A). The voltage required to obtain this switching is larger than the one at room temperature. The sample was then warmed up to 298 K, after which it was switched back to a non-metallic state (inset m in Fig. 2B) with a higher resistance of ~10 kΩ. These results are consistent with the results seen in the room-temperature device; thus, it is very difficult to attribute the nonlinear insulating properties observed in LaAlO$_3$ to the oxygen vacancy or excess oxygen scenarios.

We propose a novel mechanism to describe this nonlinear insulating behaviour. In Fig. 3A, a sketch of the band structure of the device is depicted. Our Femtosecond pump-probe experiments on single crystal LaAlO$_3$ reveal that defect states can exist in LaAlO$_3$ within a wide energy range at approximately 2.0 eV below the conduction band, which is consistent with the theoretical calculations of Luo *et al.* (*25*). Initially, there is no conduction between the Nb-SrTiO$_3$ and LaAlO$_3$. Under negative bias, charges will be injected into and fill up the LaAlO$_3$ defect levels. At the voltage where a dramatic conductivity transition is seen, the electron population of the defect levels is high enough to form a metallic state. We call this defect-mediated conduction band a "quasi-conduction band" (QCB). Once the energy level of QCB lines up (Fig. 3B) with the Fermi level of Nb-SrTiO$_3$, which is slightly above its conduction band (*26*), a continuum state is established, and this state remains even when the current is reduced to zero because of the overlap of the electron wave functions. The effective charge transferred to the defect levels at the switching threshold of 0.22 A is estimated to be around $4.6 \times 10^{19}$ cm$^{-3}$ for a device area of 0.02 cm$^2$, a LaAlO$_3$ thickness of 150 nm and a defect level lifetime of ~10 microseconds. While this carrier concentration is about 3.5 times smaller than that of the Nb-SrTiO$_3$, which is $1.6 \times 10^{20}$ cm$^{-3}$, it makes the LaAlO$_3$ a stable metallic system. To validate this idea, theoretical calculations on interstitial La$^{2+}$ defect of LaAlO$_3$ and percolation of wave functions were performed (supporting online material). It was found that the charge density higher than 0.011872 Å$^{-3}$ exists mainly at La atoms around the defect with strong *d*-orbital character, which is of the same order of magnitude as our estimated number. Interestingly, the charge density of the lowest unoccupied conduction band at Γ point around Ti atoms in cubic SrTiO$_3$ has the same magnitude as that of the defect state of LaAlO$_3$ (Fig. 3D) such that the defect La-*d* and Ti-*d* wave functions can couple if they are in proximity. The most interesting aspect of the QCB is the inherent hysteresis; the only way that this device can be restored to the original insulating state is by applying an opposing voltage, which will then remove the carriers from the QCB (Fig. 3C). Intriguingly, the transition where the system becomes an insulator (with an applied positive voltage) is close to 2.4 V, which is the difference in the bandgaps of LaAlO$_3$ and Nb-SrTiO$_3$.

To explore the dependence on the sequence of polarity switches, an initial positive voltage scan was conducted (Fig. 4A). There was no current until a threshold was reached at about 1 V, beyond which the current increased to a maximum and then decreased until the system went back to the original resistive state. This switching behaviour can be understood by considering the same defect model but with the electrons being pumped into the defect states from the valence band side. At the same time, the position of the Nb-SrTiO$_3$ conduction band is pushed down with respect to the LaAlO$_3$ band and at about 1 V the defect band has enough charge to overlap with it (to form a QCB), which leads to a stable metallic state as depicted in Fig. 4B. However, as the positive voltage is increased, the electrons in the QCB will be pumped into the conduction bands of LaAlO$_3$ and the metal electrode and the Nb-SrTiO$_3$ conduction band will be further pushed down; this leads to a lack of overlap between the bands and thus the original resistive state (Fig. 4C).

This QCB originates from the defect levels in the bandgap of LaAlO$_3$. To test this hypothesis, devices with very thin LaAlO$_3$ films (25 and 50 nm), which are likely to have fewer defects, were examined, and no switching at all was observed (supporting online material). The defect model is also validated by the change in resistive state values after several cycles; the device has slightly lower resistance values after the first cycle, and those values slightly increase in the later cycles as shown in Fig. 1, C and D. This change points to the possibility of defect production and defect rearrangement after large currents have passed through these devices, which supports the defect-mediated QCB model for nonlinear insulating $LaAlO_3$.

Most of the complex oxides have predominantly ionic bonds and are prone to a variety of cationic and anionic defects including vacancies, interstitials and antisites. These defects create a plethora of electronic states within the bandgap of these oxides and constitute a new class of materials that we have coined as "nonlinear insulators". In these insulators, the defect levels can be populated to form QCBs, which can lead to multiple conduction states in the same system. These states can be stabilised if an adjacent metallic conduction band overlaps with them. The only way to restore the levels back to their original state is by removing the carriers from the defect levels; this phenomenon is what leads to hysteresis in the *I-V* curves.

The implications of these nonlinear insulators are far-reaching. For example, the use of multi-component oxides as insulators in devices (*e.g.,* high-*k* dielectrics in silicon CMOS devices) must be exercised with caution. However, once the defect states in these materials can be controlled, researchers will be in a position to develop many novel device concepts based on these non-linear insulators.

27. The authors would like to thank Y. L. Zhao and M. Yang for experimental support and the National Research Foundation (NRF) Singapore under the Competitive Research Program "Tailoring Oxide Electronics by Atomic Control" for financial support.

**Figure legends:**

**Fig. 1. Demonstration of a nonlinear insulator**. (**A**) Sample configuration and measurement geometry. (**B**) *R-T* curve of the initial resistance state for Cu/LaAlO$_3$ (~150 nm)/Nb-SrTiO$_3$ structure on a semi-logarithmic scale. (**C**) *I-V* measurements by scanning voltage along 0 → -7 V → 0 → 7 V → 0 and *R-T* curves of different resistance states. The inset m and inset q are the *R-T* curves after scanning the voltage through points m and q to zero, respectively. The horizontal data between points k and l are due to the compliance current in action. (**D**) *I-V* curves after the cycle shown in **C** and *R-T* curves of different resistance states. The inset l and inset p are the *R-T* curves after scanning the voltage through points l and p to zero, respectively.

**Fig. 2. Low temperature switching.** (**A**) Negative switching of Au/LaAlO$_3$ (~150 nm)/Nb-SrTiO$_3$ at 4.1 K. (Inset i) *R-T* curve of the initial state. (Inset n) *R-T* curve after negatively scanning voltage back to 0, *i.e.*, after 0 → -10 V → 0 at 4.1 K. The current values in the k-l-m sequence are confined by the compliance current. (**B**) Positive scan after **A** at 298 K by 0 → 7 V → 0. (Inset) *R-T* curve after positive scan.

**Fig. 3. Schematics of QCB.** (**A**) Schematic of the band diagram of the device with no voltage bias. The middle defect band represents the defect levels of LaAlO$_3$, which are widely distributed in the bandgap at ~2 eV below the conduction band. (**B**) Formation of a QCB under an initial negative voltage bias. (**C**) Depletion of electrons in the QCB by a subsequent positive bias. (**D**) Partial charge density distributions of local defect state around the Fermi level for LaAlO$_3$ and lowest unoccupied conduction band at Γ point for cubic SrTiO$_3$, respectively. The yellow clouds represent charge densities exceeding 0.011872 electrons/Å$^3$.

**Figure 4: Switching polarity.** (**A**) *I-V* characteristics of the initial positive voltage scan for Cu/LaAlO$_3$ (~150 nm)/Nb-SrTiO$_3$ structure from 0 → 7 V → 0 following the sequence i-j-k-l-m-n. The inset shows the *R-T* curve after scanning the voltage back to zero. (**B**) Pumping of electrons from the valence band of LaAlO$_3$ by an initial positive bias greater than 1 eV. (**C**) Generation of a band offset between the QCB and the Nb-doped SrTiO$_3$ conduction band by a subsequent increase in the positive bias.

**Fig. 1., Liu *et al.***

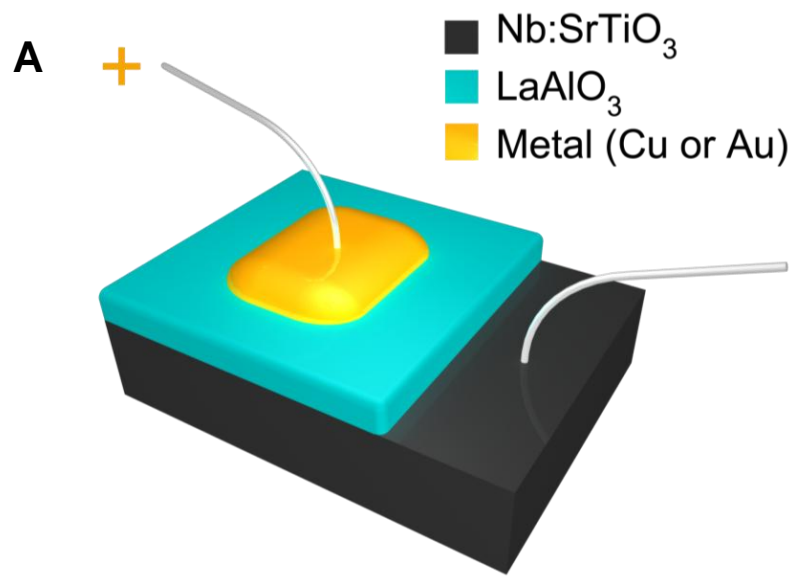

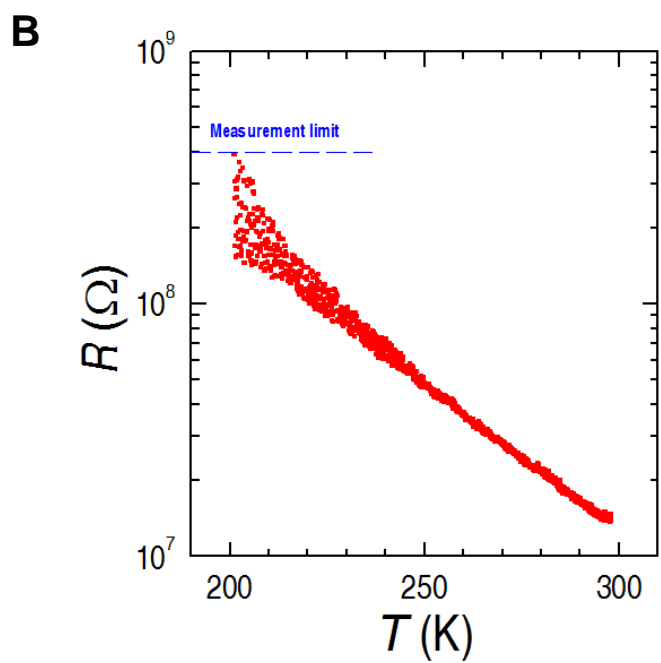

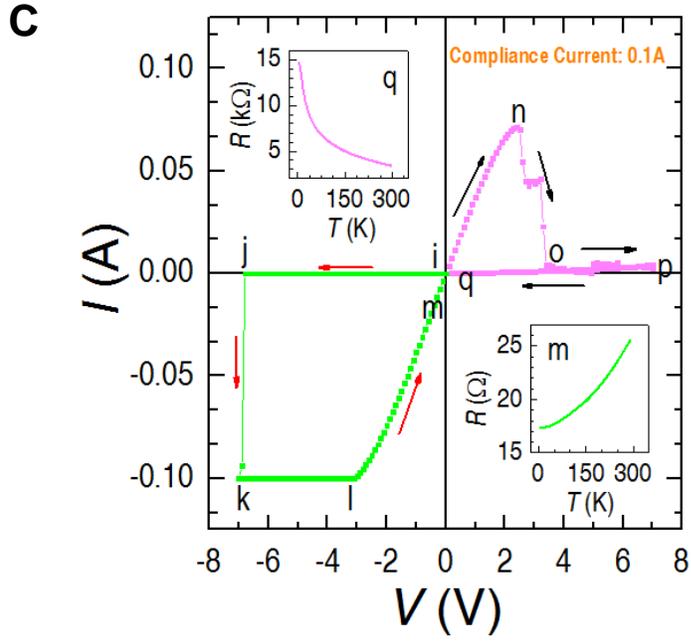

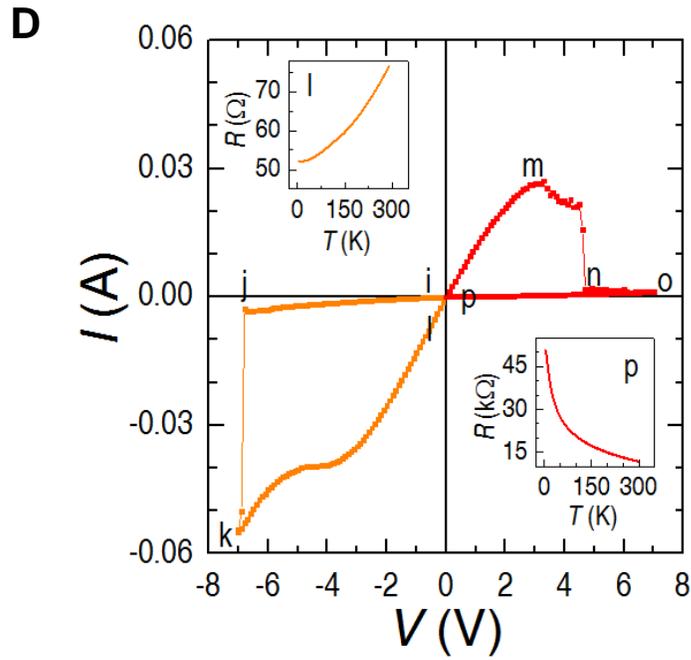

**Fig. 2., Liu** *et al.*

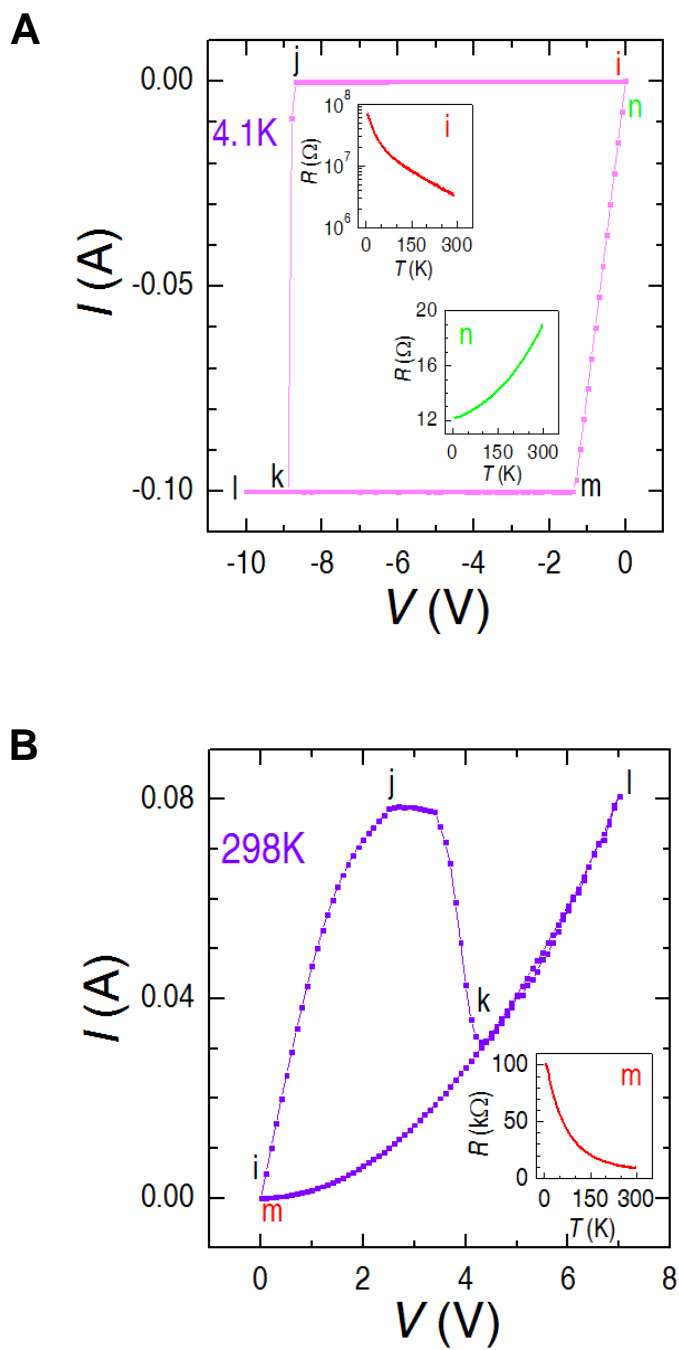

**Fig. 3., Liu** *et al.*

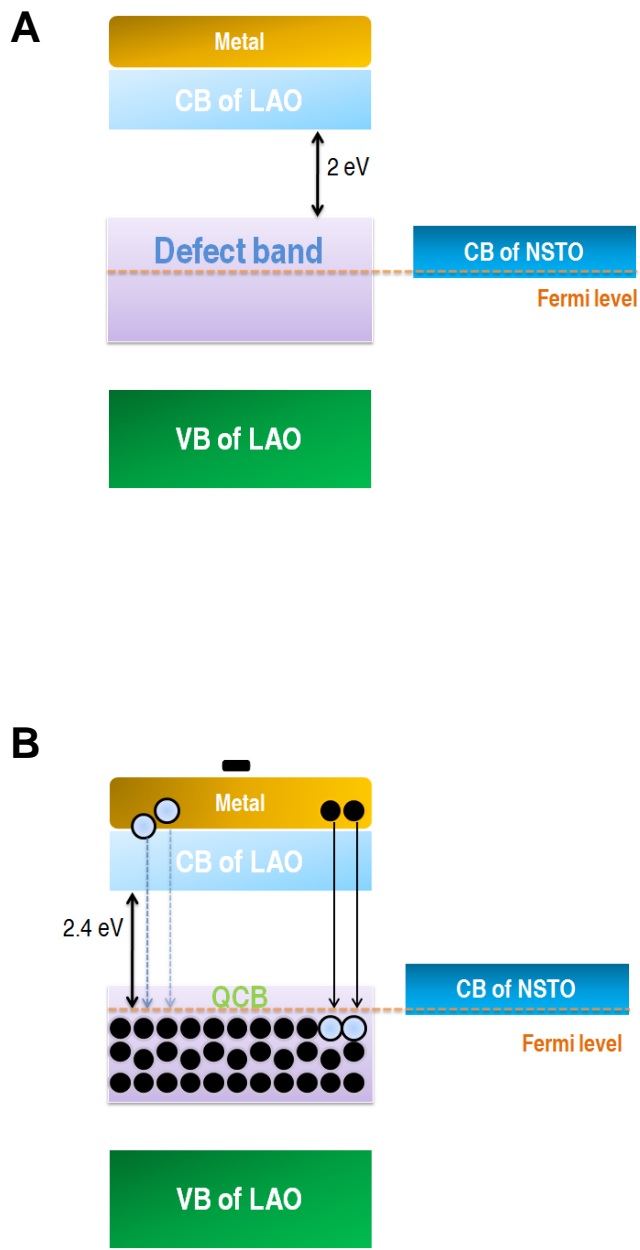

**Fig. 3., Liu *et al.*

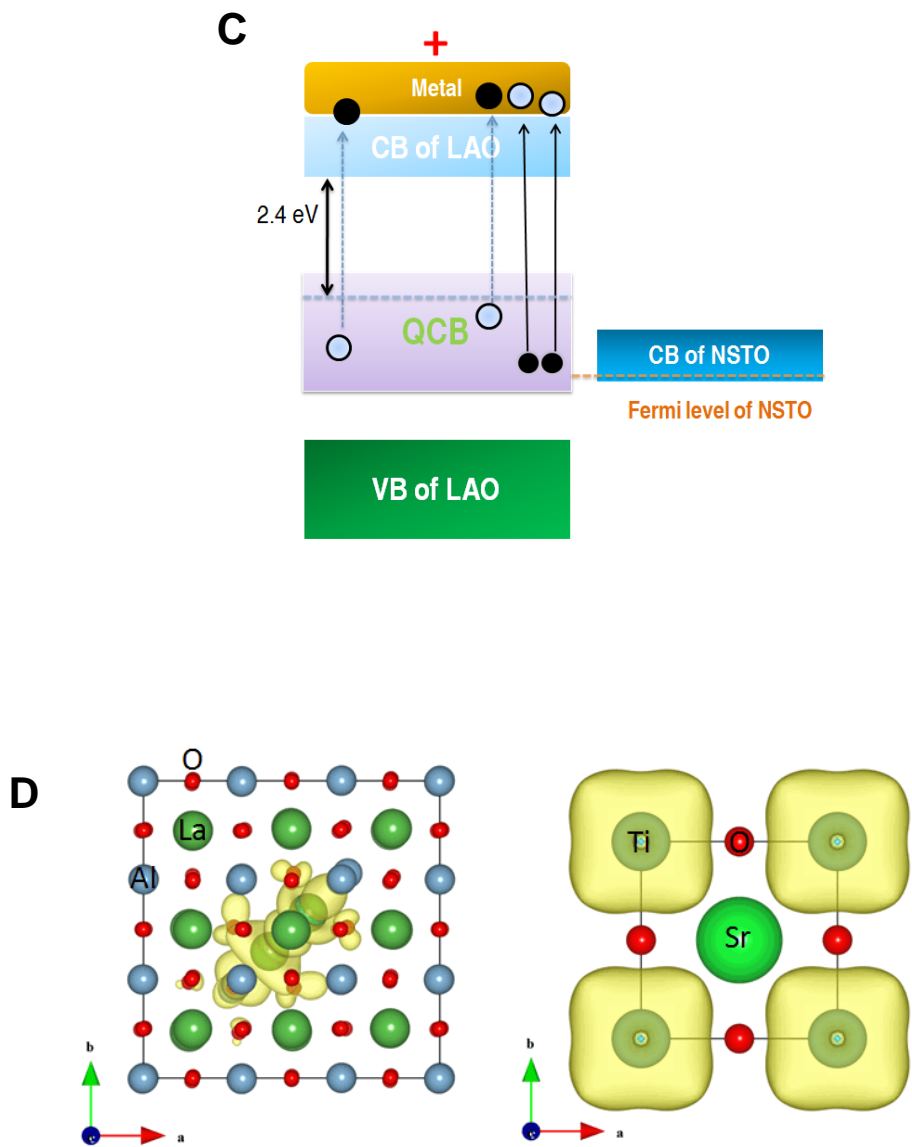

**Fig. 4., Liu *et al*.**

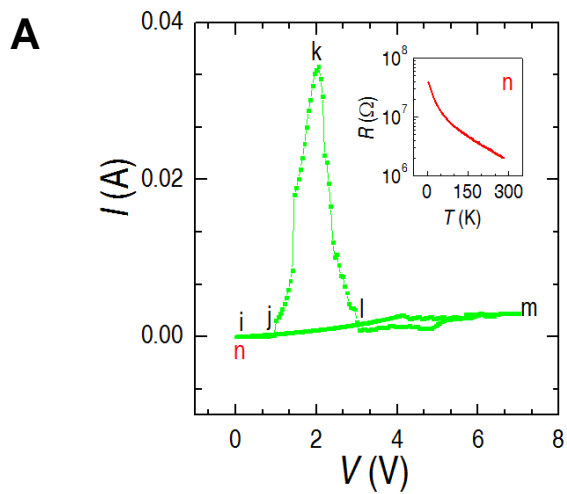

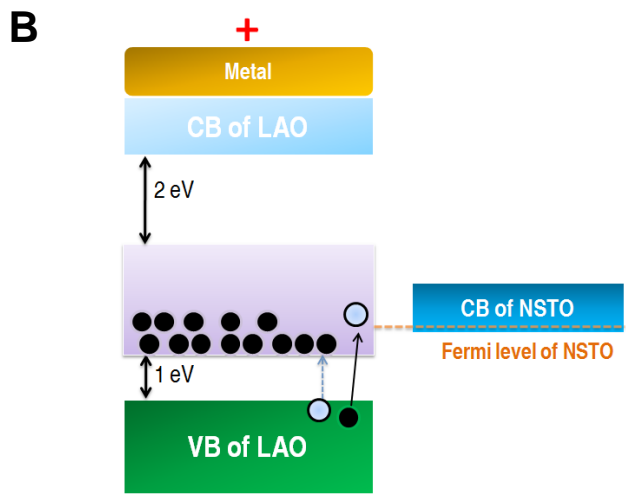

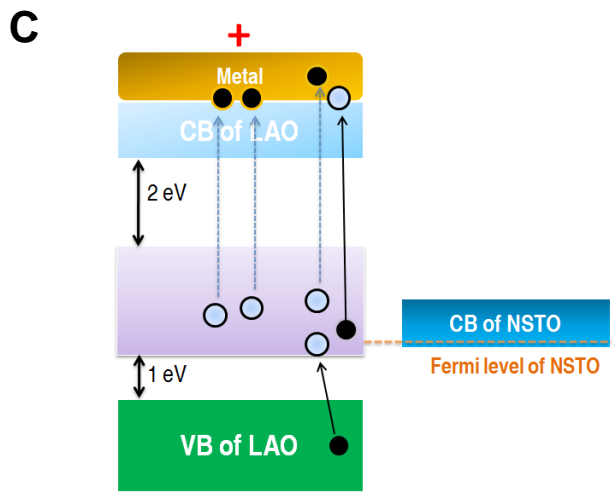